\documentclass[12pt]{article}
\usepackage[margin=1in]{geometry}
\usepackage{comment}
\usepackage{graphicx}
\usepackage{fancyhdr}
\usepackage{amsmath}
\usepackage{amssymb}
\usepackage[numbers,sort&compress]{natbib}
\usepackage{hyperref}
\usepackage{doi}

\begin{document}

\title{\textbf{Can LLMs Reason About Attention? Towards Zero-Shot Analysis of Multimodal Classroom Behavior}}

\author{
  Nolan Platt$^{1}\thanks{Incoming PhD student, University of Pennsylvania. Correspondence to: nwplatt@seas.upenn.edu}$,
  Sehrish Nizamani$^1$,
  Alp Tural$^1$,
  Elif Tural$^1$,
  Saad Nizamani$^1$, \\
  Andrew Katz$^1$,
  Yoonje Lee$^1$,
  Nada Basit$^2$ \\[6pt]
  $^1$Virginia Tech, Blacksburg, Virginia, USA \\
  $^2$University of Virginia, Charlottesville, Virginia, USA 
}

\date{}
\maketitle
\thispagestyle{fancy}

\begin{abstract}
    
  Understanding student engagement usually requires time-consuming manual observation or invasive recording that raises privacy concerns. We present a privacy-preserving pipeline that analyzes classroom videos to extract insights about student attention, without storing any identifiable footage. Our system runs on a single GPU, using OpenPose for skeletal extraction and Gaze-LLE for visual attention estimation. Original video frames are deleted immediately after pose extraction, thus only geometric coordinates (stored as JSON) are retained, ensuring compliance with FERPA. The extracted pose and gaze data is processed by QwQ-32B-Reasoning, which performs zero-shot analysis of student behavior across lecture segments. Instructors access results through a web dashboard featuring attention heatmaps and behavioral summaries. Our preliminary findings suggest that LLMs may show promise for multimodal behavior understanding, although they still struggle with spatial reasoning about classroom layouts. We discuss these limitations and outline directions for improving LLM spatial comprehension in educational analytics contexts.

\end{abstract}

\section{Introduction}

For lecturers and instructors in academia, understanding how students engage during their lectures has long been useful for improving engagement, designing new instructional methods (pedagogy), and improving classroom design. However, this has typically required manual, trained observers to sit in on lecturers, or perhaps requiring an instructor to record their lectures, only to watch them later and try to identify different markers to improve engagement. Both approaches, though, are time-consuming, expensive, and do not scale~\cite{Hur2022, Sabuncuoglu2023}. Recent automated approaches using computer vision (CV) have emerged as alternatives, but most approaches require either specific hardware (e.g., eye-tracking glasses~\cite{Alkabbany2023}), storage of identifiable video footage raising privacy concerns, or a focus on specific classification tasks without providing quantifiable insights for instructors~\cite{Lin2021, Zhang2021}.

Recent work has shown that large language models (LLMs) can analyze educational data, achieving statistically significant agreement with human observers on classroom dialogue~\cite{Long2024}. While nontrivial, these approaches are typically reliant on text transcripts rather than visual behavior data (gaze estimations, pose data, etc). Further, privacy-preserving skeleton-based approaches~\cite{Li2023} demonstrate that pose data alone can support learning analytics without retaining any identifiable imagery. This thus provides us with a compelling question: \textit{can LLMs reason about student attention and engagement directly from geometric pose and gaze data, without any task-specific training?}

In this paper, we present our preliminary research on a novel privacy-preserving pipeline that extracts skeletal poses (via OpenPose~\cite{cao2019}) and visual attention targets (via Gaze-LLE~\cite{Ryan2024}) from classroom video, immediately discards original frames, and sends the resulting geometric data to a reasoning-focused LLM (QwQ-32B-Reasoning) for zero-shot behavioral analysis. Instructors access results through a secure web dashboard displaying attention heatmaps and temporal engagement summaries. Pilot studies are underway in real classrooms at Virginia Tech with full IRB and FERPA compliance.

Our work demonstrates privacy-preserving classroom behavioral analytics, FERPA compliance, preliminary evidence on LLM capabilities for zero-shot reasoning over pose and gaze data, and design implications for AI-assisted teaching reflection tools.

\section{Related Work}

In classroom analysis, the most common CV libraries for pose detection are OpenPose~\cite{cao2019}, You Only Look Once (YOLO)~\cite{yolo}, and OpenCV~\cite{opencv}. Recent work includes a computer vision based system for engagement detection in online classrooms, focusing on facial fatigue, head focus, emotional receptivity, and behavioral involvement, which gives real time feedback to teachers and students~\cite{zhao2025}. The system focused purely on facial recognition and emotional analysis in a small, select online classroom. Li et al.~\cite{lietal} propose an AI-based system using YOLOv5 and OpenPose to analyze classroom behavior. Using a dataset of 6,687 images, the model was able to correctly recognize classroom behavior 84.23\% of the time and overall location 79.6\% of the time. The system was limited to five behaviors trained on the YOLO model: raising hand, looking down, looking up, standing up, and turning around. A recent study~\cite{impact2023} outlines the relative impact of different classroom learning behaviors (both positive and negative) through deep learning and CV. The authors observed, analyzed, and tested 180 minutes of 25 university freshmen to understand the correlation to the positive and negative behavior. Evidently, the authors showed there is statistical significance between positive behaviors and higher test scores, while negative behaviors predicted lower performance with 88\% accuracy.

Beyond pose estimation, gaze tracking has become a contender for measuring student attention in classrooms. Ryan et al.~\cite{Ryan2024} introduced Gaze-LLE, a lightweight gaze estimation model using frozen DINOv2 encoders that achieves state-of-the-art performance with only 2.8M trainable parameters. Alkabbany et al.~\cite{Alkabbany2023} deployed a multimodal engagement measurement platform that combines gaze tracking, facial expressions, and body movements in real classrooms, achieving 84\% agreement with teacher assessments. However, challenges still remain: Soares et al.~\cite{Soares2024} documented calibration drift, tracking loss from glasses reflection, and battery limitations in eye-tracking systems.

Privacy concerns (e.g., FERPA) have also motivated recent work toward skeleton-based representations. Li et al.~\cite{Li2023} proposed CVPE, a privacy-preserving system that extracts skeletal poses, immediately deletes original frames, and achieves GDPR compliance while maintaining 94\% action recognition accuracy. Carr et al.~\cite{Carr2024} showed that raw skeletonization data isn't inherently anonymous through their achievement of 68\% re-identification accuracy. They also showed adversarial motion retargeting can reduce this to 8\% while preserving behavioral information.

Large language models have also been applied to some educational contexts. Long et al.~\cite{Long2024} fine-tuned GPT-4 to analyze classroom dialogue with a Cohen's $\kappa \approx$ 0.89 across 15 behavioral categories. Kim et al.~\cite{Kim2024} introduced a real-time dashboard that processes multimodal inputs every 30 seconds to provide teaching interventions. Work by Blikstein and Worsley~\cite{Blikstein2016} showed that combining different modalities captures distinct aspects of learning, with multimodal models predicting learning gains \textit{more accurately} than single-modality approaches.

Having discussed prior work in computer vision, privacy, foundation models, gaze estimations, a clear gap can be seen. Prior systems have combined vision and language models for education, but \textbf{no prior work} combines privacy-preserving skeletonization pose extraction, gaze estimation, and zero-shot LLM reasoning to provide instructors with quantifiable, scalable, FERPA-compliant behavioral analytics without task-specific training data.

\section{System Design}

\begin{figure}[h]
  \centering
  \includegraphics[width=\linewidth]{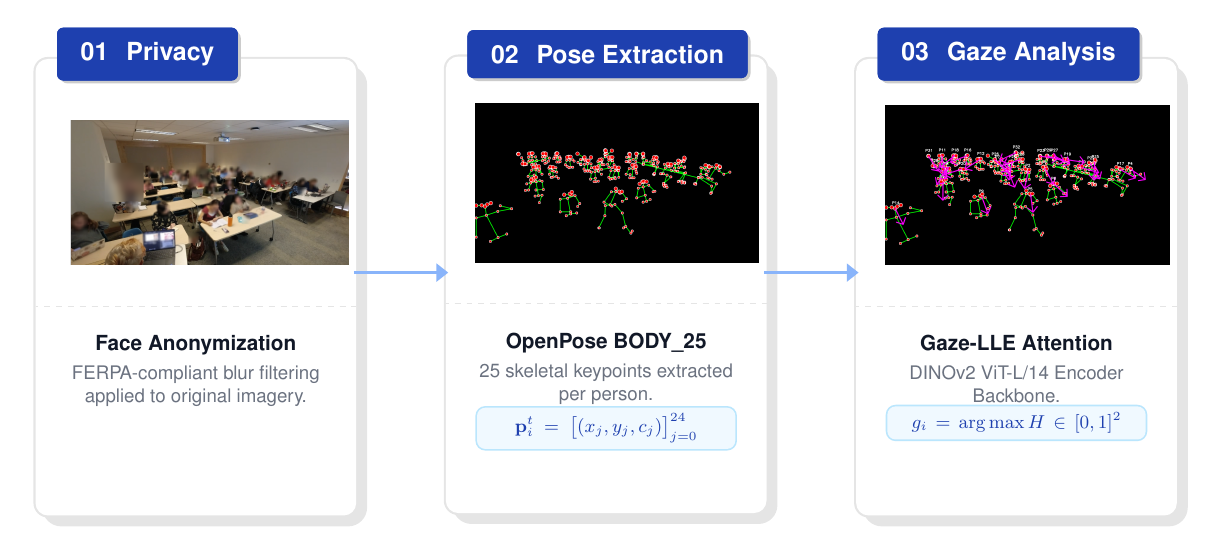}
  \caption{Privacy-preserving vision processing. Original classroom video (far left) is split into individual frames, with each frame subject to a blur filtering algorithm for face anonymization. OpenPose then extracts 25 skeletal keypoints per person (middle), and Gaze-LLE estimates visual attention vectors (right). Original frames are then permanently deleted.}
  \label{fig:stages}
\end{figure}

Our system processes classroom videos through a three-stage pipeline that prioritizes student privacy while still generating useful insights for instructors. Figure~\ref{fig:stages} shows the computer vision component of our work.

\subsection{Privacy-Preserving Vision Processing}

Face anonymization is applied \textit{immediately} upon frame extraction using Gaussian blur with a $\sigma=15$ kernel, ensuring FERPA compliance before any analysis occurs whatsoever. OpenPose~\cite{cao2019} extracts the BODY\_25 skeleton model, representing each point as a 2D keypoint with an associated confidence score.

Let $\mathbf{p}_i^t = [(x_j, y_j, c_j)]_{j=0}^{24}$ 
where $(x_j, y_j)$ are pixel coordinates and $c_j \in [0,1]$ represents detection confidence for 25 keypoints including head, torso, limbs, and extremities.

Gaze-LLE~\cite{Ryan2024} processes the same frames to estimate visual attention targets. The model uses a frozen DINOv2 ViT-L/14 encoder with a lightweight 2.8M parameter decoder, generating a spatial heatmap $H \in \mathbb{R}^{W \times H}$ from which gaze targets are extracted as $\mathbf{g}_i = \arg\max H$, normalized to $[0,1]^2$ coordinates.

Once skeletal poses and gaze vectors are extracted and saved as JSON, \textit{original video frames are permanently deleted}. This ensures that no identifiable imagery is retained in storage, maintaining FERPA compliance while preserving all behavioral information necessary for analysis.

\subsection{GPU Memory Management and LLM Processing}

The entire system is running on a single NVIDIA RTX 6000 ADA (48GB VRAM). CV libraries and LLMs demand extensive GPU memory while running. This is complemented by the specific LLM we are using (QwQ-32B-Reasoning), a reasoning-intensive model that will typically use 99\% of the GPU cores. To that end, we methodologically designed a pipeline that adhered to our specifications. The two CV libraries (OpenPose and Gaze-LLE) use $\approx$ 6GB VRAM. After both CV stages complete, we deallocate both vision models, clear the CUDA cache via \texttt{torch.cuda.empty\_cache()}, and reset memory statistics. Allocated memory is then reduced from 6GB to $\approx$ 0.01GB, freeing $\approx$ 43GB for loading QwQ-32B-Reasoning, a large reasoning model quantized to FP8 precision. To that end, the resources we are using are significantly lower than state-of-the-art approaches in computer vision; this is important to note since, as a pilot study, the computational efficiency, speed, and accuracy may only rise from here. We intentionally developed our system to be a single-GPU environment to \textbf{reflect realistic institutional constraints}.

QwQ-32B-Reasoning processes the pose and gaze data, which were saved as JSON, through a three-stage analysis that was carefully designed to stay within the model's context window. While the base model has a context window of just 8,192 tokens, we extended it to 131,072 tokens through YaRN based off recent work in language model research~\cite{platt2025, peng2023}.

First, \textbf{micro-chunk analysis} processes 60-second video segments (approximately 12 frames at 5 fps sampling). For each segment, the model is given JSON containing all student pose and gaze vector data. The model then generates detailed per-student timelines on their behavior, attention, posture, and engagement levels. After micro-chunk finishes, \textbf{synthesis aggregation} combines five micro-chunks into 5-minute summaries, identifying patterns and transitions across a longer time period. Lastly, a \textbf{final summary} processes all of the synthesis (five minute) outputs so we can see patterns and insights across the \textit{entire video}, not just specific segments.

Processing time is heavily dependent on the length of the video (and by extension, the length of the lecture). Typically, most lectures at Virginia Tech and the University of Virginia range between 60-90 minutes. From our experiments thus far, the pipeline averages approximately 2.7 hours for a one-hour classroom recording: vision processing requires 30 minutes, while LLM analysis (60 micro-chunks + 12 syntheses + 1 final summary) requires approximately 140 minutes of compute time. The processing time per video can be significantly reduced in the future through potential grants, funding, and compute access; the main blocker we have is only possessing a singular GPU. While more than capable of running the entire pipeline, it was a nontrivial disadvantage that mandated meticulous design of the pipeline.

\subsection{Web Dashboard and Instructor Interface}

\begin{figure}[h]
  \centering
  \includegraphics[width=0.85\linewidth]{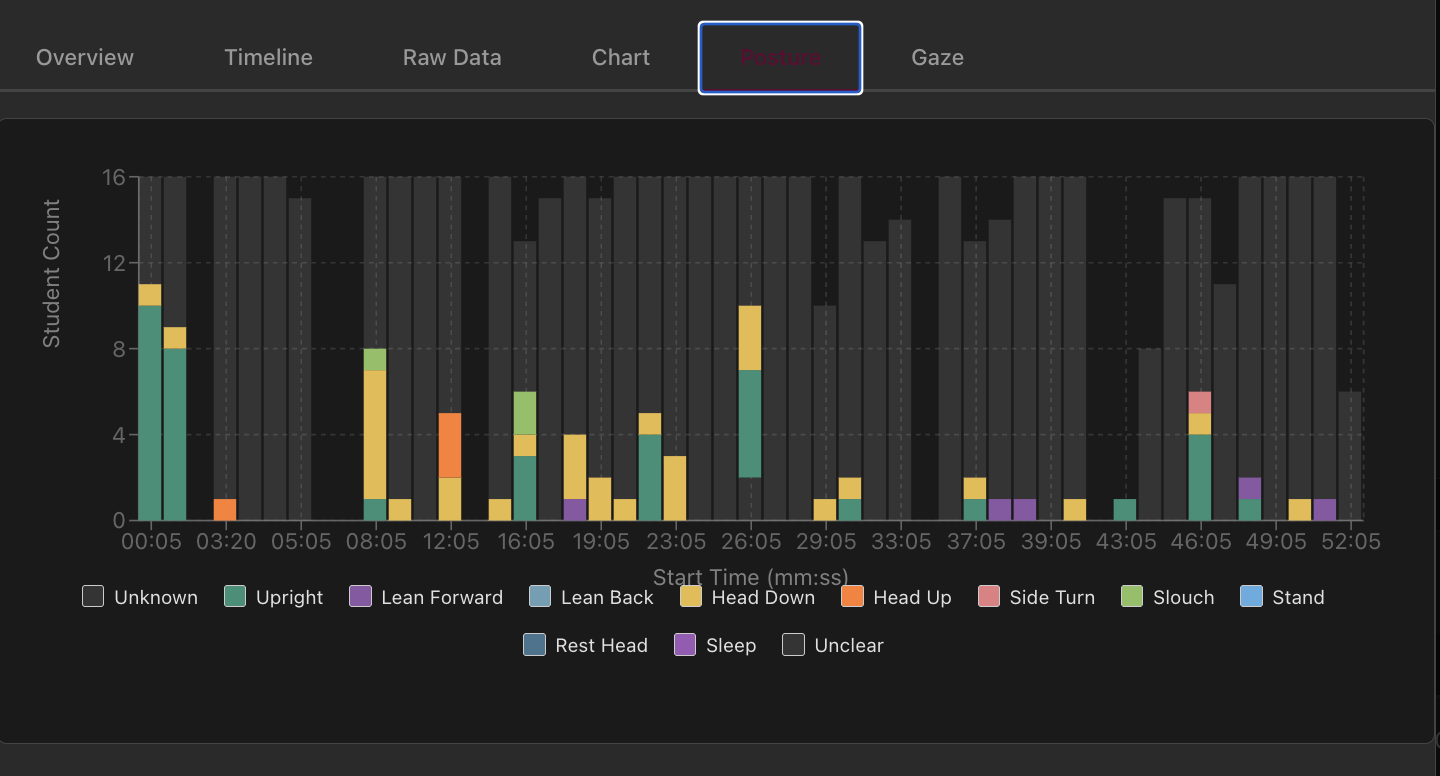}
  \caption{A bar graph showing representative change in posture throughout a lecture's video, showing periods of time where students leaned forward, were sleeping, slouching, standing, etc. Unknown classifications include occlusion and low-confidence keypoints. High unknown rates are a known limitation that is being addressed.}
  \label{fig:dashboard}
\end{figure}

All instructors participating in the pilot study can upload videos, track pipeline progress and ETA of completion, and access the corresponding results through a web dashboard. Using a FastAPI backend, the dashboard aggregates data and generates spatial attention heatmaps that show gaze targets across the classroom (or lecture hall for larger areas), per-student engagement timelines showing posture quality and visual focus over time, and LLM-generated behavioral summaries with frame-level citations for verification. The system uses Celery task queues for asynchronous processing and WebSocket connections for real-time progress updates during analysis. We designed the dashboard to be highly scalable, hoping this pilot study will eventually expand to more departments, universities, and instructors. The Celery task queue, for instance, was designed in such a way that dozens of videos could be uploaded on a single GPU -- but it will only process \textit{one video at a time} before moving onto the next, ensuring we do not exceed our computational limits. Figure~\ref{fig:dashboard} shows an example of the temporal engagement visualization available to instructors on the dashboard.

\section{Preliminary Findings}

  We take note of the following three distinct takeaways from our pilot study thus far, both from instructor feedback and analysis of the system as a whole:
\begin{enumerate}
    \item \textbf{Spatial Attention Patterns are Interpretable} \\ We have demonstrated that our system successfully identifies distinct gaze distributions, with clear differences between lecture-focused periods and independent work (i.e., note taking or working on a laptop). However, instructors report that the heatmaps do show clear "dead zones" in various classroom layouts, where some students' pose or gaze data are misinterpreted by the system or sometimes completely missed.
    \item \textbf{Engagement Transitions are Captured} \\ LLM analysis can find behavioral shifts with a reasonable level of accuracy. Qualitative comparison of recorded videos with LLM-generated analysis does suggest the system captures behavioral shifts. A formal, IRB-approved, validation study with $\approx$15 human evaluators using Cohen's $\kappa$ is currently underway.
    \item \textbf{LLMs Still Struggle with Spatial Reasoning} \\ Perhaps the most significant limitation of our research, unsurprisingly, involves complex spatial relationships. While it would be arbitrary to state that \textit{all} LLMs continue to struggle with spatial reasoning, it is fair to say that QwQ-32B-Reasoning struggles to understand highly complex classroom settings. For example, it consistently struggles to understand that students looking "left" could be viewing a secondary screen rather than being distracted, leading to misclassified attention events. 
\end{enumerate}

\section{Discussion}
Our preliminary work builds the groundwork for LLM-based classroom analytics, including our revelation of both the promise and current limitations of such systems. Rather than overwhelming instructors with random metrics, the visualizations and heatmaps in the dashboard provide instructors with more actionable decision points for pedagogical reflection. Nonetheless, the current limitations with LLM's ability to perform spatial reasoning present an unanswered problem in applying foundation models to physical domains. LLMs are excellent at temporal pattern recognition but continue to lack the spatial comprehension necessary for complex classroom understanding. In future work, we will investigate several different paths to address this issue, including a custom model context protocol (MCP) with tool use for spatial grounding of different classrooms (e.g, providing the board, screen, and door coordinates as structured context). We will also incorporate a vector database for handling context across different segments, and perhaps converting classroom layouts to raw JSON data for semantic understanding.

\section{Conclusion and Future Work}

We have presented a privacy-preserving classroom engagement analysis system that combines computer vision, LLM reasoning, and interface design. We have shown that our approach provides meaningful insights for instructors, derived directly from geometric pose and gaze data while maintaining FERPA compliance. The spatial reasoning limitations we identify underline directions for improving LLM spatial comprehension in physical contexts. Future work will address computational requirements, spatial reasoning enhancements, and longitudinal impact studies.

\bibliographystyle{unsrtnat}
\bibliography{references}

\end{document}